\documentclass[12pt,preprint]{aastex}

\newcommand{\Mjup}{M$_{\rm Jup}$}
\newcommand{\Lsun}{L$_{\sun}$}
\newcommand{\Rsun}{R$_{\sun}$}
\newcommand{\BD}{CFHT-BD-Tau~4}

\shorttitle{Detailed Look at a Brown Dwarf Disk}
\shortauthors{Pascucci et al.}

\begin{document}


\title{The First Detailed Look at a Brown Dwarf Disk}


\author{I. Pascucci, D. Apai, Th. Henning}
\affil{Max--Planck--Institut f\"{u}r Astronomie,
       K\"{o}nigstuhl 17, D-69117 Heidelberg, Germany}
\email{pascucci@mpia.de, apai@mpia.de, henning@mpia.de}

\and
\author{C. P. Dullemond} 
\affil{Max--Planck--Institut f\"{u}r Astrophysik, 
	      Karl--Schwarzschild--Str. 1, D-85741 Garching, Germany}
\email{dullemon@MPA-Garching.MPG.DE}



 
\begin{abstract}
The combination of mid-infrared and recent submm/mm measurements allows us
to set up the first comprehensive spectral energy distribution (SED)
of the circumstellar material around a young Brown Dwarf.
Simple arguments suggest that the dust is distributed in the form of a disk. 
We compare basic models to explore the disk parameters.
The modeling shows that a flat disk geometry fits well the observations.
A flared disk explains the SED {\it only} if it has a puffed-up inner rim and
an inner gap much larger than the dust sublimation radius.
Similarities and differences with disks around T~Tauri stars are discussed.

\end{abstract}




\section{Introduction}

 Hundreds of brown dwarfs (BDs) have been discovered in the last few
years, both within star-forming regions \citep{mor03,Persi} 
and among field stars \citep{phan,Kirk}. 
Still very little is known about their formation processes.

First indications that BDs may form as ordinary stars, i.e. via accretion from a 
circumstellar disk, come from observations of H$\alpha$ \citep{Jay03} 
and near-infrared excess emission \citep{mue01,liu03}. 
However, in the case of BDs  only excess emission
at longer wavelengths provides the indisputable evidence for the presence
of circumstellar material \citep{NattaTesti}. 
Thermal emission of warm dust around BDs has been observed  by a number of
groups \citep{Comeron,Persi,testi01,apai}.
Very recently, our search for submm/mm emission from young
BDs provided the first detection of cold dust around two of them
\citep{klein}.
For the first time, this allowed to estimate the mass of the circumstellar
material and to relate it to the mass and luminosity of the central object. 
 
These results open up new interesting questions:
{\it Is the circumstellar material around BDs confined in a disk? Are these disks
similar to T~Tauri disks or different ones?
What is the effect of the ultra-low luminosity on the disk properties?}
Tentative modeling of disks has been performed by \cite{NattaTesti} based on two
ISOCAM mid-infrared (MIR) measurements with the general assumption of flared disk geometry.
The first ground-based observations of one of their objects showed that the flared disk
model cannot be applied blindly: at least in one case a flat configuration
can better explain the observations \citep{apai}. 
The same flat geometry has been later used by \cite{Natta}
to interpret the ISOCAM measurements of other BDs.       
 
A thorough investigation of the disk properties requires measurements of
different disk zones, i.e. in different wavelength domains.
In this Letter we compile the first complete spectral energy distribution (SED) of a disk 
around a BD.
The goal of this paper is to test the basic models and determine
 which model parameters can be constrained from the
combination of short- and long-wavelengths observations.
Our study focuses on the object \BD{} 
(2MASSI J0439474+260140)
which was found and confirmed as a young BD by \cite{mar01}. 
Its very strong H$\alpha$ emission \citep{mar01} coupled with X-ray activity
\citep{mok} and further spectral signatures \citep{Jay03} prove that this BD is
intensively accreting matter.

In the following, we first present the SED of \BD , we prove that the circumstellar
material cannot be in a spherical shell and we argue towards a disk configuration.
We model the SED applying different disk geometries.
Finally, we discuss similarities and differences with accretion disks around
T~Tauri stars.
 
\section{The Spectral Energy Distribution}

In this section we compile the observations available on \BD .
Optical and near-infrared data, together with an estimate of the 
line-of-sight extinction (A$_{\rm V}$=3~mag), have been published by \cite{mar01}.
An additional L-band flux is given in \cite{liu03}.   
We retrieved archival ISOCAM images at 6.7 and 14.3~\micron{} and we measured
fluxes of 28.4 and 38.5~mJy by using an aperture of 24\arcsec .
We accept a photometric accuracy of 20\% as given in the
ISOCAM Calibration Accuracies 
Document\footnote{http://www.iso.vilspa.esa.es/manuals/CAM/accuracies/CAM\_accur.pdf}.
Our group observed \BD{} in the mm and submm continuum at
the IRAM-30m and the JCMT telescopes.
The data reduction and analysis is presented in \cite{klein}.
We provide the summary of the above fluxes and errors in Table~\ref{flux}.
         
\section{Input Parameters}

The main input parameters for the modeling are derived directly from the
observations.
\BD{} is expected to be a member of the Taurus star-forming region located at a distance
of 140 pc.
We assume an effective temperature of $2900 \pm 100$~K for an M7$\pm$0.5 spectral type
\citep{mar01,briceno} and a mass of 70~\Mjup . 
In agreement with \cite{mar01}, we calculate a
total luminosity of 0.1 \Lsun{}  which is brighter than the luminosity of 
the youngest (1 Myr) isochrone from \cite{allard}.

The wavelength dependence of the dust opacity ($\kappa_{\lambda}$) and 
the mass of the circumstellar dust can be calculated from the expected 
optically thin submm and mm fluxes.
In the optically thin regime the flux depends on the wavelengths as 
F$_{\nu} \propto \lambda^{-(2+\beta)}$, where
$\beta$ is the wavelength dependence of $\kappa_{\lambda}$. 
A value of $\beta =$~2  has been measured for 
interstellar dust particles \citep{dr84}, while $\beta =$~1  is generally 
used for circumstellar disks where grain growth is expected to take place \citep{beck00}.
In our case the slope of the SED is $-3.8\pm0.9$, with the error computed
from the uncertainty in the flux measurements.
This implies that $\beta$ is $1.8\pm0.9$.
Due to the large error on the measured $\beta$, we cannot 
fix the dust opacity. To compare our
results with disks around T~Tauri stars, we assume the same dust opacity as suggested
by \cite{beck90} which amounts to 2~cm$^{2}$/g at 1.3~mm.
For a dust temperature between 10 and 20~K and a gas-to-dust ratio of 100,
we calculate a total circumstellar mass of 1.5--0.3~\Mjup .
We note that adopting the 10 times lower opacity of interstellar grains 
would lead to 10 times larger circumstellar masses.

\section{Modeling}
 
A simple spherical shell model is first investigated by using the 
radiative transfer (RT) code of \cite{manske}. 
The mass of the circumstellar material is set to 4~\Mjup{}
in order to reproduce the submm and mm fluxes. Different power-law density
distributions, inner/outer radii and dust properties were tested (see
Table~\ref{Models}). 
The 1D simulations always predict a visual extinction at least an order 
of magnitude higher than measured by \cite{mar01} and \cite{it02}.
Thus, the simple shell configuration has to be excluded and a 2D distribution has
to be considered.
   
The most plausible 2D circumstellar dust distribution is believed to be in the
form of a disk \citep{te93}.
In the following, we test various versions of the most widely accepted disk
configurations, i.e. the flat \citep{adams} and the
flared disk \citep{kl87}.
The opening angle of the flat disk is constant, while that of the 
flared disk increases with radius. An optically thin superheated layer
 is expected for flared configurations.
An addition to some of the flared disks is the self-consistent treatment of the inner rim that,
due to its higher temperature, puffs-up with respect to the flaring disk behind it \citep{kees01}.
Similar changes in the flared disk geometry have been proposed by \cite{cg97}
as an effect of stellar accretion and viscous dissipation.
For all the models, we adopt a dust opacity which follows the silicate mass absorption
coefficient by \cite{dr84} for short wavelengths and the opacity law by \cite{beck90}
 for long wavelengths. The transition wavelength depends on the grain size: for small grains of
 0.1~\micron{} radius the two opacity laws meet at 100~\micron , while for
 10~\micron{} grain they meet at 200~\micron .

We consider two flat disk models (m4 and m5 in Table~\ref{Models}): 
one with a single disk temperature distribution
\citep{beck90} and one with an additional hot surface layer. 
The main difference between the two approximations is at 10~\micron , where
the two-layer model predicts an emission feature, which is absent in
the optically thick one-layer flat disk.
For both configurations, plausible physical parameters provide a good fit.  
Because of the lack of measurements at the silicate feature,
we cannot discriminate between the two cases.

\begin{figure} 
\plotone{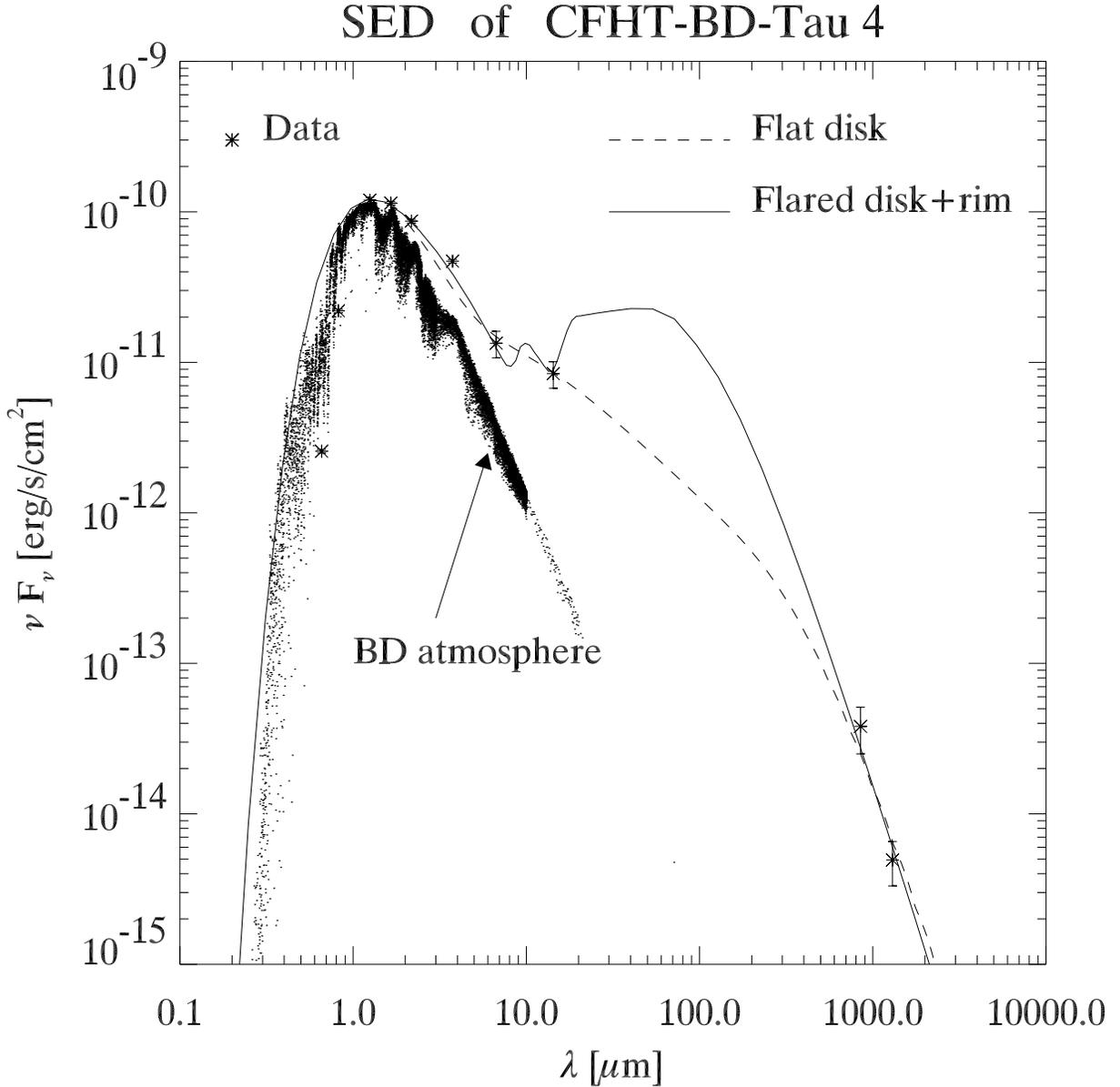}
\caption{Comparison of model fits with observations.
The measurements are de-reddened with a visual extinction of 3 mag \citep{mar01}
and are plotted with asterisks. The photometric errors are taken from  Table~\ref{flux}.
The models that best fit the observations are the flat disk 
(dashed line, model m4 in Table~\ref{Models})
and the flared disk with an inner puffed-up rim 
(solid line, model m9 in Table~\ref{Models}).
The atmosphere of a 1~Myr old BD with 3000~K effective temperature
is overplotted in dots \citep{allard}. \BD{} is more luminous than the youngest isochrone from
\cite{allard}.\label{sed}}
\end{figure}

For the flared disk configurations \citep{kees01}, a too high MIR flux is predicted 
by the models when the disk inner radius is determined by the dust sublimation temperature.
To reduce the MIR emission three possibilities are considered: high 
disk inclination, a large inner gap and the shadow of a puffed-up inner 
rim.
Considering the low line-of-sight extinction measured towards \BD , almost
edge-on disks (inclinations larger than 60\degr) are not plausible. 
Even when a large inner gap of about 0.07~AU  is introduced in the
classical flared disk model (m7), the MIR measurements at 6.7 and 14.3~\micron{}
cannot be simultaneously fitted.
On the contrary, the flared disk with a puffed-up inner rim and large inner gap
reproduces well the SED of \BD{} (m9 in Table~\ref{Models} and solid line in Fig.~\ref{sed}). 

To demonstrate the effect of grain growth, we also tested an opacity law for silicate 
grains with 10~\micron{} radius \citep{dr84}.
Such grains lack the emission feature at $\lambda=10$~\micron . 
The one-layer flat disk model is not affected by the different dust opacities.
For the two-layer flat configuration, the silicate feature at 10~\micron{} vanishes, but the
SED can still be well fitted with parameters similar to those provided in Table~\ref{Models}.
Because larger grains have lower opacities in the MIR regime, the flared disk model with the 
puffed-up rim can fit the observations with a slightly smaller inner gap (0.04~AU). 
Even with the addition of larger grains, the classical flared disk model
does not explain the observed SED.

\section{Discussion} 

The combination of short and long wavelengths observations 
allow us, for the first time, to investigate the properties 
of the circumstellar dust around a BD. 
The SED shows an overall shape similar to that of T~Tauri stars.
Adopting the opacity by \cite{beck90}, we calculate a circumstellar mass
of  0.3--1.5~\Mjup , which is a few percent of the BD's mass.
A similar mass ratio is found for T~Tauri stars and their disks.
Thus, the disk is unlikely to be truncated by any previous ejection event
\citep{Reipurth}. A discussion is presented in \cite{klein}.

Since spherically distributed dust would result in too high extinction,
we argue towards the presence of a circumstellar disk.
Our disk models suggest that {\it either the disk is flat, or it is
flared with a puffed-up inner rim and a large inner gap}.
A flat geometry was previously suggested 
by \cite{apai} to explain the SED of a disk around a BD candidate. 
The present results support a flat disk as well, thus 
suggesting that this  might be general for disks around very 
low-luminosity objects. A flat dust distribution could be a hint towards 
large grain sizes since large grains are expected to settle in the
disk midplane (see \cite{kl87} and \cite{cg97} for a discussion of the physical 
conditions under which flared disks can exist). 
Future observations at 10~\micron{} could further strengthen 
this picture if no silicate feature is found. 

Should the disk be flared with a puffed-up inner rim, an inner
gap 5 times larger than the dust sublimation radius is required to
account for the observed SED.
The presence of magnetic fields is not sufficient to explain inner gaps 
as large as those found in the modeling (about 9~\Rsun), see for example \cite{konigl}.
The material close to the BD could be cleared out by a massive planet,
as suggested in the case of  TW Hya by \cite{calvet}.
However, the most realistic explanation is the presence of a close companion with 
a spectral type similar to the primary. 
This  would also explain the over-luminosity of this BD \citep{mar01}.
Unfortunately, the estimated separation of the possible companion would be $< 1$~mas, 
unresolvable with any current imaging technique.

To fully characterize BD disks, a detailed exploration of their
SED is essential. Especially important are the silicate emission features and
the wavelength range between 60-180~\micron. The upcoming SIRTF and SOFIA missions
will be capable of gathering a statistically relevant number of complete SEDs, as
well as providing a reliable disk frequency.

\section{Conclusions}

We present the first comprehensive models of the circumstellar
dust surrounding a young BD.
Our results are:
\begin{itemize}
  \item the overall shape of the SED is very similar to those of T~Tauri disks
  \item a spherical dust distribution contradicts the observed extinction
  \item a simple flat disk fits well the observations
  \item the flat geometry hints towards large grains 
  \item the classical flared disk does not reproduce the SED of \BD
  \item a flared disk with a puffed-up inner rim can only explain the observed SED 
    if it has an inner gap much larger than the dust sublimation radius 
\end{itemize}

\clearpage

\begin{deluxetable}{c c c c}
\tablewidth{0pt}
\tablecaption{Observed fluxes for CFHT-BDTau 4\label{flux}}           
\tablehead{            
\colhead{$\lambda$}  & \colhead{F$_{\nu}$}  & \colhead{Error}   & \colhead{Reference}\\  
\colhead{[\micron ]}  & \colhead{[mJy]}     &   \colhead{[mJy]} & }	    
\startdata
0.658              & 0.062                    & 0.002             & 1 \\
0.822              & 1.23                     & 0.03              & 1 \\ 
1.25               & 22.9                     & 1.0                & 1 \\
1.65               & 38.7                     & 1.0                & 1 \\
2.17               & 46.5                     & 1.4                & 1 \\
3.75               & 52.5                     & 1.5                & 2 \\ 
6.7                & 28.4                     & 5.7               & 3 \\
14.3               & 38.5                     & 7.7               & 3 \\
850                & 10.8                     & 3.7\tablenotemark{a} &4 \\
1300               & 2.14                     & 0.7\tablenotemark{a} &4 \\
\enddata
\tablenotetext{a}{The large uncertainty in the submm and mm measurements
is due to the uncertain flux of the calibrators. The estimated statistical errors 
are smaller (1.8 and 0.56~mJy, respectively).}
\tablerefs{(1) \cite{mar01};(2) \cite{liu03};(3) This paper;(4) \cite{klein}}
\end{deluxetable}

\clearpage

\begin{deluxetable}{lcccccccc}
\tablewidth{0pt}
\tablecaption{Summary of the Model Parameters and Results }
\tablehead{
\colhead{ID}        &
\colhead{Type\tablenotemark{a}}          & \colhead{Geom.}  &
\colhead{Param.\tablenotemark{b}}          & \colhead{R$_{\rm in}$}    &
 \colhead{Fits?\tablenotemark{c}}  & \colhead{M$_{\rm disk}$} &
\colhead{Comments\tablenotemark{d}} & \colhead{Refs.} \\
\colhead{ }        &
\colhead{ }        & \colhead{ }  &
\colhead{ }        & \colhead{[AU]}    &
\colhead{ }        & \colhead{[\Mjup]} &
\colhead{ }        & \colhead{ } 
}
\startdata

m1& RT & Shell & 0   &       & No & 4&A$_{\rm V} \sim$60~mag& (1)   \\
m2& RT & Shell & $-$1 &      & No & 4&A$_{\rm V} \sim$300~mag& (1)   \\
m3& RT & Shell & $-$2 &      & No & 4&A$_{\rm V} \sim$1000~mag& (1)  \\
m4& AA & Flat  & one-layer   & 0.02   & Yes &  1.4&  $i$=20--45\degr(35\degr)  & (2)  \\
m5& AA & Flat  & two-layer   & 0.01   & Yes &  4&  $i$=5--30\degr(10\degr)  &    \\
m6& SCC& Flared&No Rim& 0.03 & (Yes) &  0.7& $i$=80-82\degr & (3) \\
m7& SCC& Flared&No Rim& 0.07 & No & 0.7 &  & (3) \\
m8& SCC& Flared& Rim  & 0.03 & No & 0.3 & & (3) \\
m9& SCC& Flared& Rim  & 0.05 & Yes  & 0.3&$i$=8--20\degr(12\degr) & (3) \\
\hline
\enddata
\tablecomments{Summary of the models and results. Given are the models' identifier, 
the geometry (shell or disk type), the characteristic parameters, inner disk radius,
possibility of fitting the observations, total disk mass, 
comments and references  
\label{Models}} 


\tablenotetext{a}{RT - Radiative Transfer Code; AA - Analytical Approximation;
SCC - Self-Consistent Calculation}
\tablenotetext{b}{For 1D models we provide the exponent of the power-law density distribution}
\tablenotetext{c}{A parenthesis stands for models which are physically not
		plausible}
\tablenotetext{d}{A$_{\rm V}$ stands for visual extinction, $i$ is the disk inclination (0\degr for face-on)
		with the best disk inclination in parenthesis}
\tablerefs{(1) \cite{manske};(2) \cite{beck90};(3) \cite{kees01}}

\end{deluxetable}

\clearpage

\clearpage

\acknowledgments
We thank Wolfgang Brandner for helpful discussions.
This work is based on ISO data archive.

\clearpage



\clearpage

\end{document}